# Complex structures of dense lithium: electronic origin

**V F Degtyareva**

Institute of Solid State Physics Russian Academy of Sciences, Chernogolovka, Russia

E-mail: degtyar@issp.ac.ru

**Abstract.** Lithium - the lightest alkali metal - exhibits unexpected structures and electronic behaviour at high pressures. As the heavier alkalis, Li is *bcc* at ambient pressure and transforms first to *fcc* (at 7.5 GPa). The *post-fcc* high-pressure form Li-*cI*16 (at 40-60 GPa) is similar to Na-*cI*16 and related to more complex structures of heavy alkalis Rb-*oC*52 and Cs-*oC*84. The other high pressure phases for Li (*oC*88, *oC*40, *oC*24) found at pressures up to 130 GPa are specific the only to Li. The different route of Li high-pressure structures correlates with its special electronic configuration containing the only 3 electrons (at 1*s* and 2*s* levels). Crystal structures for Li are analyzed within the model of Fermi sphere - Brillouin zone interactions. Stability of *post-fcc* structures for Li can be supported by Hume-Rothery arguments when new Brillouin zone plains appear close to the Fermi level producing pseudogaps near the Fermi level and decreasing the crystal energy. The filling of Brillouin-Jones zones by electron states for a given structure defines the physical properties as optical reflectivity, electrical resistivity and superconductivity. To understand complexity of structural and physical properties of Li above 60 GPa is necessary to assume the valence electrons band overlap with the upper core electrons and increase the valence electron count under compression.

## 1. Introduction

Recent high pressure studies of the structure and properties of elements under high pressure bring considerable revision to what has been known previously (see review papers [1-3] and references therein). At normal conditions, the group-I elements of the Periodic table from Li to Cs at normal conditions are related to the free electron metals and adopt the body-centered cubic (*bcc*) structure. They transform under pressure to the face-centered cubic (*fcc*) structure and at significant compression to open and complex structures. Li is the lightest metal containing the only 3 electrons (at 1*s* and 2*s* levels) and its special electronic configuration results in a different sequence of Li high-pressure structures. For dense lithium a departure from nearly free-electron behaviour and tendency to "a pairing of the ions" has been predicted by Neaton and Ashcroft [4]. The *post-fcc* high-pressure form Li-*cI*16 (at 40-60 GPa) [5] is similar to Na-*cI*16 and related to more complex structures of heavy alkalis Rb-*oC*52 and Cs-*oC*84. The other high pressure phases for Li (*oC*88, *oC*40, *oC*24) found at pressures up to 130 GPa are specific the only to Li [6,7].

Physical properties of Li under compression change essentially. An increase in resistivity has been reported under dynamic and static pressure [8, 9]. A remarkable increase of the superconducting temperature from 0.4 millikelvin at ambient pressure to 14 – 20 K at pressures 40 – 70 GPa [9-11]. Recently a reentrant transition semiconductor to metal was found for Li at 80 – 120 GPa [12]. Unusual drop was observed in the melting temperature of lithium below 200 K at ~40 GPa [6].

For heavier alkali metals all changes in structure and properties usually are accounted by the conversion of valence electrons from $s$ to $d$ states [13], with the upper empty $d$-band moving downward on compression and overlapping with the $s$-band. For lithium a pressure driven $s - p$ transition is considered [5]. It is assumed that at high degrees of volume compression the valence electron are localized in the interstitial sites forming a '*pseudebinary*' ionic or '*electride*' compound [14].

In this paper possible causesare investigated that contribute to the formation of the complex crystal structures in lithium under pressure and compare these with structures of heavier alkalis. The electronic energy contribution is suggested as a cause for the formation of complex crystal structures and changes the physical properties.

## 2. Theoretical background and method of analysis

The crystal structure of metallic phases is defined by two main energy contributions: electrostatic (Ewald) energy and the electron band structure term. The latter usually favours the formation of superlattices and distorted structures. The energy of valence electrons is decreased due to a formation of Brillouin planes with a wave vector q near the Fermi level $k_F$ and opening of the energy pseudogap on these planes if $q_{hkl} \approx 2k_F$ [15]. Within a nearly free-electron model the Fermi sphere radius is defined as $k_F = (3\pi^2 z/V)^{1/3}$, where z is the number of valence electrons per atom and V is the atomic volume. This effect, known as the Hume-Rothery mechanism (or electron concentration rule), was applied to account for the formation and stability of the intermetallic phases in binary simple metal systems like Cu-Zn, and then extended and widely used to explain the stability of complex phases in various systems, from elemental metals to intermetallics [16-19].

The stability of high-pressure phases in lithium are analyzed using a computer program BRIZ [20] that has been recently developed to construct Brillouin zones or extended Brillouin-Jones zones (BZ) and to inscribe a Fermi sphere (FS) with the free-electron radius $k_F$. The resulting BZ polyhedron consists of numerous planes with relatively strong diffraction factor and accommodates well the FS. The volume of BZ's and Fermi spheres can be calculated within this program. The BZ filling by the electron states ($V_{FS}/V_{BZ}$) is estimated by the program, which is important for the understanding of electronic properties and stability of the given phase. For a classical Hume-Rothery phase $Cu_5Zn_8$, the BZ filling by electron states is equal to 93%, and is around this number for many other phases stabilized by the Hume-Rothery mechanism.

Diffraction patterns of these phases have a group of strong reflections with their $q_{hkl}$ lying near $2k_F$ and the BZ planes corresponding to these $q_{hkl}$ form a polyhedron that is very close to the FS. The FS would intersect the BZ planes if its radius, $k_F$, is slightly larger then $q_{hkl}/2$, and the program BRIZ can visualize this intersection. One should keep in mind that in reality the FS would usually be deformed due to the BZ-FS interaction and partially involved inside the BZ. The ratio $2k_F/q_{hkl}$, called a "truncation" factor or a "closeness" factor, is an important characteristic for a phase stabilized due to a Hume-Rothery mechanism. For Hume-Rothery phases such as $Cu_5Zn_8$, this closeness factor is equal 1.015, and it can reach up to 1.05 in some other phases. This means that the FS radius can be up to approximately 5% larger than the BZ vector $q_{hkl}/2$ for the phase to be stabilized due to a Hume-Rothery mechanism. Thus, with the BRIZ program one can obtain the qualitative picture and some quantitative characteristics on how a structure matches the criteria of the Hume-Rothery mechanism.

## 3. Results and discussion

Several high-pressure structures of lithium are analyzed to reveal effects of the Hume-Rothery mechanism on occurrence of structural complexity. The Li-*cI*16 structure is considered as a distortion and superlattice of the *bcc* structure. The Li phase with the orthorhombic 24-atom cell is compared with one of the phases in Na (structural data are presented in table 1).

**Table 1.** Structure parameters of Li phases as given in the literature. Fermi sphere radius $k_F$, the ratio of $2k_F$ to Brillouin zone vectors ($2k_F/q_{hkl}$) and the filling degree of Brillouin zones by electron states $V_{FS}/V_{BZ}$ are calculated by the program BRIZ [20].

| Phase | Li-*bcc* | Li-*cI*16 | Li-*oP*24 | Na-*oP*8 |
|---|---|---|---|---|
| | | *Structural data* | | |
| Space group | $Im\bar{3}m$ | $I\bar{4}3d$ | *Pbca* | *Pnma* |
| P,T conditions | Ambient conditions | P=39 GPa | P=115 GPa | P=119 GPa |
| Lattice parameters (Å) | $a$ = 3.5091 | $a$ = 5.2716 | $a$ = 4.213 $b$ = 4.205 $c$ = 7.482 | $a$ = 4.765 $b$ = 3.020 $c$ = 5.251 |
| Atomic volume Å$^3$) | 21.605 | 9.156 | 5.52 | 9.45 |
| $V/V_0$ | 1 | 0.424 | 0.255 | 0.239 |
| Reference | [1] | [5] | [6,7] | [21] |
| | | *FS - BZ data from the BRIZ program* | | |
| $z$ (number of valence electrons per atom) | 1 | 1 | 2 | 2 |
| $k_F$ (Å$^{-1}$) | 1.110 | 1.479 | 2.205 | 1.844 |
| Total number of BZ planes | 12 | 24 | 50 | 30 |
| $k_F /(½ q_{hkl})$ max min | 0.877 | 1.013 | 1.055 0.875 | 1.074 0.886 |
| Filling of BZ with electron states $V_{FS}/V_{BZ}$ (%) | 0.5 | 0.89 | 0.932 | 0.932 |

### 3.1. Li-*cI*16 structure as the Hume-Rothery phase

The *post-fcc* phase of Li at pressures ~39 GPa has a rhombohedral cell and is specific to lithium [5]. The structure Li-*hR*1 is a distortion of the fcc structure, with the compression along the body-diagonal and an increase of the rhombohedral angle from 60° to 62.87° (*c/a* decreases from 2.45 for *fcc* to *c/a* = 2.296 for Li-*hR*1 at 39.8GPa). At the *fcc* − *hR*1 transition the (111) *fcc* plains divided in two sets: one set (101) is approaching closer to the Fermi sphere and another set (003) is moving away. Closer contact of (101) plains to the FS leads to the gain in the band structure energy.

The first complex phase of lithium was found at pressures above 40 GPa with the *cI*16 structure [5]. Li-*cI*16 is 2 × 2 × 2 superlattice of *bcc* with slight shifts of atoms resulting in the formation of new (211) planes just in contact with the FS. The new BZ is ~90% filled by the electron states, satisfying the Hume-Rothery rules [2, 16].

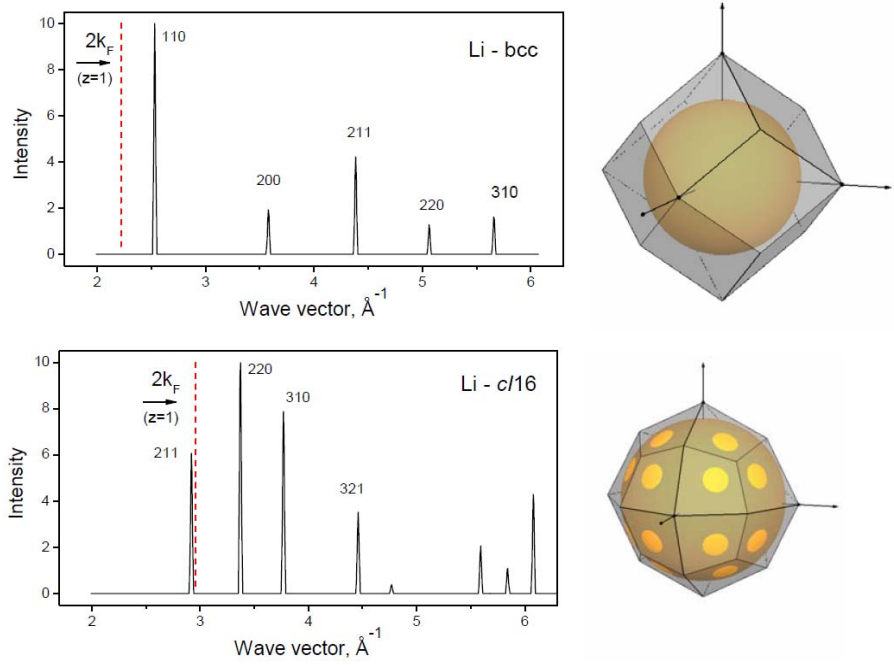

**Figure 1.** Calculated diffraction patterns (left) and corresponding Brillouin zones with the inscribed FS (right). In the top panel – for Li-*bcc*, *cI*2 and in the bottom panel – for Li- *cI*16 with structural parameters given in table 1. The position of $2k_F$ for given $z$ and the *hkl* indices are indicated on the diffraction patterns. Brillouin zones are shown in common view with *a*\* pointing forward and *c*\* pointing up.

### 3.2. Valence electron count in Li above 60 GPa

The Li structure found at pressures above 95 GPa was identified as an orthorhombic cell containing 24 atoms [6, 7]. Several space groups were suggested for this structure [7]. For our analysis the *oP*24 type structure was selected with the experimentally found lattice parameters [6] as given in table 1. The Li-*oP*24 structure can be considered in relation to the *oP*8 structure found for heavier alkali metals Na and K [21, 22]. The diffraction patterns of Li-*oP*24 and Na-*oP*8 in figure 2 are characterized by the appearance of a group of reflections close to $2k_F$ and hence by a formation of several Brillouin zone planes near the FS. Configurations of BZ's are similar for both phases with some additional planes for Li-*oP*24 due to formation of superlattice along one axis. The prototype hexagonal cell is NiAs-*hP*4 type and its distortion to *oP*8-type was analyzed in the case of Na [16]. Cell parameters *a*, *b*, *c* of the orthorhombic phases are related to the hexagonal cell parameters $a_h$, $c_h$, as follows:

$$c_h, a_h, a_h\sqrt{3} \quad \text{for Na-}oP8 \quad \text{and} \quad a_h\sqrt{3}, c_h, 3a_h, \quad \text{for Li-}oP24.$$

Valence electron number $z = 2$ is assumed for Li-*oP*24 as in Na-*oP*8 and AuGa-*oP*8 phases [16]. The double ionic radius for coordination number CN = 8 - 11 for lithium was estimated by Shannon [23]. These values are $0.92\text{Å} \times 2 = 1.84\text{Å}$ and $1.1\text{Å} \times 2 = 2.2\text{Å}$ for CN equal 8 and 11 respectively. The interatomic distances for Li-*oP*24 structure are very close to these values implying core–valence band overlap at such compression. For structure stability it is necessary to assume core ionization: an overlap of the core electrons with the valence electron band of *sp* type. Resulting valence electron count should be increased and univalent alkali metal Li turns into a divalent metal. Similar analysis has been performed for Na and it has been suggested to become a divalent metal in the *oP*8 phase [20]. For *post–cI*16 phases of Li with compression $V/V_o$ < 0.35 an overlap of valence and core electron levels was suggested with an increase of effective valence electron number. This allows an understanding of the semiconducting properties of *oC*88 and *oC*40 phases as filling of BZ by electron states and returning Li to metal in the *oP*24 phase.The BZ filling by electron states for Li-*oP*24 is ~93% (see table 1) − the typical value for the metallic brass-type phases [16].

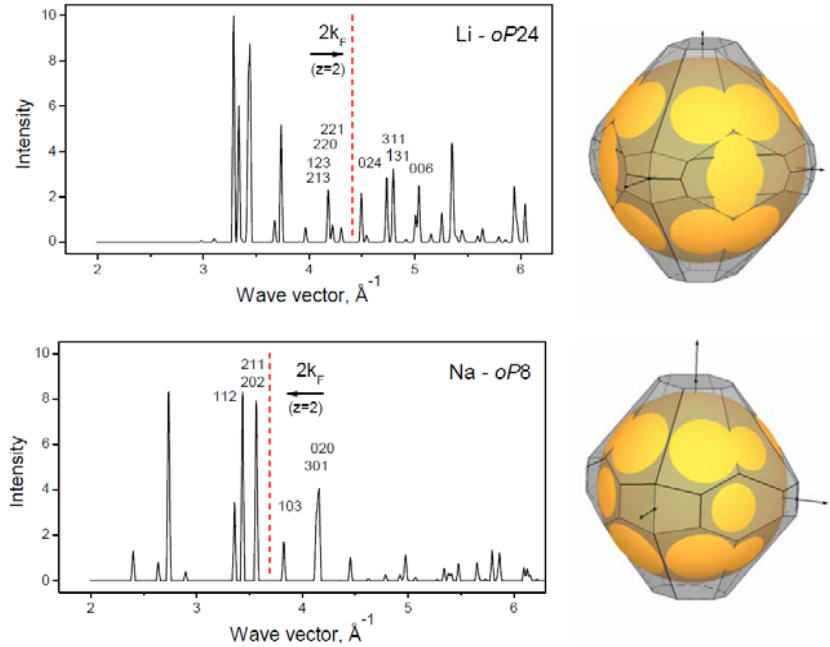

**Figure 2.** Calculated diffraction patterns (left) and corresponding Brillouin zones with the inscribed FS (right). In the top panel – for Li-*oP*24 and in the bottom panel – for Na-*oP*8 with structural parameters given in table 1. The position of $2k_F$ for given $z$ and the *hkl* indices of the principal planes are indicated on the diffraction patterns. Brillouin zones are shown for Li-*oP*24 in common view and for Na-*oP*8 in the view with *c** pointing forward and *b** pointing up.

For the heavier alkali metals Na and K along with *oP*8 phases incommensurate host-guest structures were observed [1,3]. Tsese structures were discussed in [24] with the suggestion of an overlap of the valence band and the upper core electrons resulting in the higher valence electron counts. For the heaviest alkalis K, Rb and Cs a complex structure *oC*16 has been found, the same as for group-IV (Si, Ge) elements. These findings support an assumption for heavy alkalis an increase of valence electron counts to four due to core ionization [17].

**Summary**

The formation of the Li-*cI*16 structure at pressures around 40 GPa is explained with the increase of the Hume-Rothery mechanism under compression. The ambient pressure *bcc* structure undergoes distortions and superlattice formation leading to the creation of a new 24-planes Brillouin zone that accommodates the Fermi sphere with the ~90% filling by electron states. It is considered a close structural relationship of the alkali metal structures Li-*oP*24, Na-*oP*8 and K-*oP*8 with the binary alloy phase AuGa-*oP*8. The latter phase is related to the family of the Hume-Rothery phases that is stabilized by the Fermi sphere–Brillouin zone interaction where a decrease in the electronic band structure energy occurs due to the contact of the Fermi sphere and Brillouin zone planes. An important characteristic is degree of Brillouin zones filling by electron states which depends on the count of valence electrons per atom and the number of atoms in the cell.

From similar considerations for alkali elements a necessary condition for structural stability emerges in which the valence electron *sp* band overlaps with the core electrons and the valence electron count increases under compression. Consideration of the core–valence electron band transfer may promote a better understanding of non-traditional behaviour of alkali elements under significant compression. Changes of physical properties of Li under pressure can be accounted for with the increase of valence electron energy contribution and moreover with the overlap of core and valence electron bands. Within applied model main tendencies can be understood for structural and physical transformations in elements under strong compression.

**Acknowledgments**

The author thanks Dr Olga Degtyareva for valuable discussions. This work was partially supported by the Program of the Russian Academy of Sciences "The Matter under High Pressure".